\documentclass[conference]{IEEEtran}
 
\usepackage[dvipsnames]{xcolor}
\usepackage{tikz}
\usepackage{hyperref}
\usepackage{multirow}
\usepackage{cite}
\usepackage{amsmath}
\usepackage{comment}
\usepackage{colortbl}
\usepackage{multirow} 
\usepackage{array}

\ifCLASSINFOpdf
\else
\fi
\usepackage{url}

\usepackage[]{fancyhdr} %
\newcommand{\changefont}{\fontsize{9}{9}\selectfont}
\IEEEoverridecommandlockouts
\begin{document}

%
\title{Locational Aspect of Fast Frequency Reserves in Low-Inertia Systems - Control Performance Analysis}

\author{\IEEEauthorblockN{Georgios Misyris, Deepak Ramasubramanian, Parag Mitra, Vikas Singhvi}
\IEEEauthorblockA{Grid Operations \& Planning Group, \\Electric Power Research Institute, \\
Knoxville, Tennessee, USA\\
GMisyris@epri.com}}
\maketitle
\thispagestyle{fancy}
\pagestyle{fancy}


\begin{abstract}
This paper evaluates the frequency performance of an AC system when primary frequency response is provided by inverter-based resources located at remote-areas. Due to potentially larger wave propagation constants over longer lines, fast active power response from inverter based resources may have a negative impact on the system frequency response. Within this context, this paper presents a control performance analysis is presented in order to identify limitations for improving the frequency stability when inverter-based resources in remote locations use local frequency measurements. Our results suggest that there exists a trafeoff between disturbance rejection and stability robustness when allocating primary frequency control. In particular, fast frequency control can have a negative impact on the damping ratio of poorly damped electromechanical modes.
\end{abstract}



%
\IEEEpeerreviewmaketitle

\section{Introduction}
The replacement of synchronous generators with power electronics-interfaced devices decreases the available inertia in the system and can lead to much faster frequency dynamics in the grid \cite{milano2018foundations,ulbig2014impact,Misyris2018}. As a result, the dynamic behavior can endanger the system by stressing the control and protection schemes \cite{schiffer2019online}, e.g. (i) when frequency drops by unacceptable amounts, it results in the disconnection of production units and loads and producing a cascade effect that can lead to widespread power outages \cite{Winter2015} and (ii) rate of change of frequency based loss-of-main protection and under-frequency load shedding \cite{Cao2016}.

There are various ways to mitigate the impact of reduced level of inertia on the frequency dynamics, with the main focus being on utilizing the fast action of inverter-based resources. This is realized by adding a frequency controller in the outer control loops of power electronic devices. Various frequency control schemes have been proposed, such as virtual inertia and droop-based frequency control \cite{poolla2019placement,jiang2020dynamic,misyris2019grid,obradovicfrequency}. To evaluate their impact on frequency performance of the system after a large disturbance, system operators mainly use the maximum frequency deviation and the rate of change of frequency of the Center Of Inertia (COI) frequency \cite{entsoe,agneholm2019fcr,Obradovic2018,tosatto2021towards}. One common assumption that system planners make in these studies is that nodal frequencies are equal to the center of inertia frequency. This assumption simplifies the control design and the process of deciding the amount of fast frequency reserves. However, as generation sources interconnect to weak areas of the network and move farther away, the local frequency measurements might differ from the center of inertia frequency, since inertia are inhomogeneously distributed across the grid. Consequently, it is necessary to evaluate the frequency performance in low-inertia power grids and the ability of inverter-based resources to provide fast and secure primary frequency control.


Metrics such as $H_2$-norms have been used in order to evaluate the frequency performance of the system without the need for time-domain analysis \cite{Paganini,gross2017increasing,jiang2017performance}. However, in the aforementioned studies the impact of the location of the active power regulation units have been neglected. As presented in \cite{Ramasubramanian2021}, weak electrical ties, i.e. long transmission lines, prevent the delivery of frequency response from adjoining areas, thus inverter-based resources cannot improve the frequency performance of the system. 


To this end, this paper is to evaluate how much frequency droop response over long transmission lines is needed for improving the frequency stability for a power system with low inertia. Motivated by ongoing research on how inertia location and slow network modes determine how disturbances propagate in power grids \cite{pagnier2019inertia}, we try to tackle the problem of how fast frequency droop response affects the dynamic performance of the system. Within this concept, this paper provides a framework for evaluating the robust performance of inverter-based resources frequency controllers. The disturbance response ratio is chosen as a performance metric that allow to evaluate the impact of parameters on the local and center of inertia frequency dynamics. The contributions of this paper are summarized as follows. A theoretical background that enables to analyze the robust performance of frequency controllers in low-inertia systems with active power regulating units located in remote areas. Insights are provided concerning the control design of frequency droop-based controllers. Time-domain studies are performed to further analyze locational impact of delivering fast frequency response.

The reminder of this paper is organized as follows. First, a motivation is given to highlight the problems that might arise when delivering primary frequency response over long transmission paths. Then, a framework is presented that allows to analyze the frequency performance of the system. Time-domain studies are performed to further analyze locational impact of delivering fast frequency response.

\section{Problem formulation}
In larger power networks, along long transmission lines, electromechanical waves have a finite and discernible propagation time. Consider an electrical event such as a change in generation/load. If this change were to occur at one end of a long tranmission line, then the magnitude of this disturbance seen at the other end is not only lower but it is alose seen later in time \cite{Ramasubramanian2021,pagnier2019inertia}. This behavior is described as weak electrical link and could possibly have a large impact on the frequency response of a power system dominated by inverter-based resources. In this section, we present the theory that will be used in this paper for analyzing the robust performance of the system when primary frequency response is provided by inverter-based resources located far from the disturbance location.




\subsection{Control performance analysis}
\begin{figure}[t]
\centering
\includegraphics[width=1\linewidth, trim=3cm 7cm 4cm 10cm,clip]{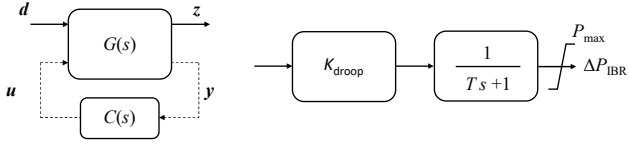}
\caption{Left figure: Block diagram for general control configuration, Right figure: droop-based inverter-based controller.}
\label{fig:Control_diagram}
\end{figure}


Fig. \ref{fig:Control_diagram} shows the closed system configuration, where $G(s)$ is the transfer function mapping the plant input $u$ to the plant outputs $y$ and performance variables $z$. The transfer function $C(s)$ is the system controller described as \mbox{$u=-C(s)y$}. The performance variables $z$ specify the desired characteristics of the closed loop system and play crucial role on the control design. 

If the performance variables $z$ are also the measured outputs $y$, the stability robustness of the system can be analyzed as follows. The transfer function $G(s)C(s)=L(s)$ is the open loop gain, which can be used to investigate how a sinusoidal signal propagates through the feedback loop. One could analyze the stability of the system by evaluating the Nyquist plot of the transfer function $L(s)$. To ensure that the system is stable, one needs to ensure that the Nyquist curve of $L(s)$ not to encircle the point (-1,0) \cite{skogestad2007multivariable}. This is done by appropriately designing the plant controller C(s). The further the Nyquist curve is from the (-1,0) point, the more stable the system is. A way to ensure that r is sufficiently large is by checking the maximum value of the magnitude of the sensitivity function, i.e. $|S|$. The sensitivity function is \mbox{defined as}:
\begin{equation}
    S = \frac{1}{1+L(s)}
\end{equation}
and indicates how a signal affects the output of the system. It holds that the sensitivity matrix should be 1 when $s\rightarrow \infty$ and 0 when $s\rightarrow0$. Based on the selection of the control gain and the bandwidth of the controller a reduction of the sensitivity function can be achieved at some angular frequencies, but this has to be compensated by amplification in other frequencies. Small $|S|$ is desirable at low frequencies in order to minimize the impact of disturbances on the steady state error. 

The above procedure for achieving robust performance of the closed system is followed whenever the performance variables $z$ are similar to the measured outputs $y$. However, in most applications the performance variables differ from the measured outputs.

\subsubsection*{Application to frequency stability}
To monitor the power imbalances in the system and reserve a sufficient amount of frequency response, the center of inertia frequency of the system has been used as a metric, i.e. the center of inertia frequency is considered as the performance variable z of the system. The COI frequency is calculated as the inertia weighted (where all inertias are on a common system base) sum of all generator frequencies in the system and the basis for this is outlined in and the expression for $N$ generators is given below: 
\begin{equation}
    f_{\rm COI} = \frac{\sum_{i=1}^N H_i f_i}{\sum_{i=1}^N H_i}
\end{equation}
where $H_i$ is the inertia constant of the generator and $f_i$ is the generator frequency. One common assumption that system planners make in order to decide the frequency response of the system, is that center of inertia frequency, which is the performance variable can be considered as a measured output $y = f_{\rm COI}$. This means that the nodal frequencies are equal to the center of inertia frequency. By making this assumption, robust frequency performance can be achieved by setting constraints on the sensitivity function $S$. This approach is followed by Nordic system operators in Europe for ensuring robust frequency performance in the system \cite{entsoe} . 

This simplifies the control design of $C(s)$ (see Fig. \ref{fig:Control_diagram}), which represents the required frequency response in order to ensure frequency stability of the system. However, as inertia level decreases and active power regulating units are located in remote areas, the nodal frequencies differ from the global one. This introduces control limitations for improving the frequency performance of the system, since the performance variables differ from the measured outputs. Consequently, a different control performance analysis is required for assessing the frequency performance of the system and designing fast primary frequency control.

\subsection{Disturbance response ratio}
To analyze the ability of a controller to attenuate external disturbance, the following transfer function can be defined as a metric:
\begin{equation}
    \centering
    T_{zd}(s) = G_{zd}(s) - G_{zu}(s)C(s)\Big (1+G_{yu}(s)C(s)\Big)^{-1}G_{yd}(s)
\end{equation}
where $T_{zd}(s)$ is the closed-loop system from the disturbance d to the performance variable $z$. The transfer function $G_{zd}(s)$ is the open-loop system from the disturbance d to the performance variable $z$. It should be mentioned that the performance variable does not have to be the same with the measured output of the system. In case the performance variable is the system with the measured variable it holds that $G_{zd}(s)$=$G_{yd}(s)$. The transfer functions $G_{zu}(s)$ and $G_{yu}(s)$ show the relationships from the disturbance d to the performance variable $z$ and measured output $y$, respectively. The transfer function C(s) is the plant controller, which takes as an input the measured output $y$ and gives as an output the input to the plant $u$. 

Good disturbance attenuation can be achieved if the amplitude of the closed-loop system $T_{zd}(s)$ is smaller compared to the amplitude the open-loop one $G_{zd}(s)$. This means that the controller K(s) should be designed so that $|T_{zd}(s)|<|G_{zd}(s)|$. How well the controller is able to attenuate an external disturbance applied to the system depends on the bandwidth of the controller but more importantly on the measured output that is used as a feedback signal.  Multiplying the $T_{zd}(s)$ with $G_{zd}(s)^{-1}$ yields the following expression:
\begin{equation}
\centering
    R_{zd}(s) = 1 - G_{zu}(s)C(s)\Big(1+G_{yu}(s)C(s)\Big)^{-1}G_{yd}(s)G_{zd}(s)^{-1}
\end{equation}
which is defined as the disturbance response ratio and it can be used as a metric to quantify the ability of the controller to attenuate disturbances. The aim is to design the feedback controller so that $|R_{zd}(s)|<1$ \cite{bjork2019influence}. To limit the effect of external disturbances $\boldsymbol{d}$ to the performance variables $\boldsymbol{z}$, we need to increase to increase the volume of the region in which $|R_{zd}(s)|<1$. By doing that, we can limit the impact of the disturbance on the frequency dynamics of the system, i.e. reducing the maximum frequency deviation. Moreover, the disturbance response ratio acts as a replacement of the sensitivity function for improving the damping of the system. By constraining the peak of $|R_{zd}|$, one can limit the disturbance amplification over the whole frequency range and improve the damping of the system.

\subsection{State space representation}
Power system dynamics can be described by a set of non-linear differential and algebraic equations. Since a power system is a highly non-linear dynamical system, to perform the study a linearization of the system needs to be performed. After the linearization, the system takes the following state space form:
\begin{align}
\centering
 \boldsymbol{\dot{x}}&=\boldsymbol{A} \boldsymbol{ x}+\boldsymbol{B_u}  \boldsymbol{  u}+\boldsymbol{B_d}  \boldsymbol{  d} \nonumber\\
 \boldsymbol{ y}&=\boldsymbol{C_y}  \boldsymbol{ x}+\boldsymbol{D_{yu}} \boldsymbol{  u}+\boldsymbol{D_{yd}}  \boldsymbol{  d} \\
 \boldsymbol{ z} &=\boldsymbol{C_z} \boldsymbol{ x}   \nonumber 
\end{align}
where $x$ is the state vector containing the state variables of the system, u is a vector containing the controlled input variables, $d$ contains the external disturbances to the system, y contains the measured outputs that can be used as feedback signals and $z$ contains the performance variables that can be used to evaluate the control and system performance. Using Laplace transformation, the system can be expressed in frequency domain, in order to use the theoretical background presented above. Figure~\ref{fig:Control_diagram} shows the block diagram for general control configuration. $G(s)$ is the system expressed in frequency domain and $C(s)$ is the plant controller, which for this report is considered as a frequency droop-based controller that can be placed in different locations in the system. In the context of this report, the input vector $u$ contains the active power outputs of the units that participate in frequency regulation, the output vector $y$ contains the measured quantities used by the active power regulation units, such as the local frequencies, $d$ is a vector of active power imbalances (modeled as step changes) that can occur at different locations in the network and the vector $z$ contains the center of inertia frequency. From a physical standpoint, an unexpected active power imbalance (step change of $d$), will result in variation of the local and center of inertia frequencies both during and steady state conditions. The control should be designed such that: (i) the maximum and steady-state frequency deviations are limited and (ii) excitation of local and inter-area modes is avoided. This means that the controller should be designed, in order to reduce the impact of external disturbances to the system.

\section{Limitations Imposed by Local-Frequency measurements \& Time Domain Analysis}
This section focuses on the analysis of the control limitations for improving the overall frequency stability, when using local-frequency measurements.
\begin{figure}[t]
\centering
\includegraphics[width=1\linewidth, trim=1.5cm 4cm 7.5cm 3cm,clip]{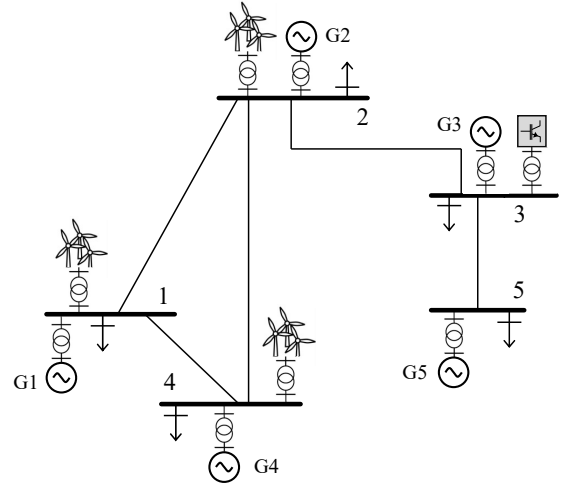}
\caption{Test system. Base case scenario with the total kinetic energy in the network being equal to 110 GWs.}
\label{fig:TestSystem}
\end{figure}
To illustrate the potential problems that may arise, the test system depicted in Fig.~\ref{fig:TestSystem} is utilized. More details about the system parameters can be found in \cite{bjork2021dynamic}. Two type of synchronous generators are assumed, hydro (located at bus 1, 2 and 3) and thermal generators (located at bus 4 and 5). All generators are equipped with automatic voltage regulators and power system stabilizers. It is considered that only hydro generation units are equipped with governor models. Moreover, inverter-based units are also equipped with frequency controllers, which can be used for providing frequency response. Since the objective here is to evaluate the impact of inverter-based resources (IBR)s on frequency response, it should be mentioned that the inner loop dynamics of the inverter-based resources are neglected as it is assumed that these control loops are tuned and designed appropriately. Table~\ref{T:1} depicts the amount of kinetic energy at each bus in the system and Table~\ref{T:2} the length of the lines.


\begin{table}[t]
\centering
\caption{Kinetic energy and active power output of generators for the high- and low-inertia scenarios.}
\begin{tabular}{c|c c|| c c}
 \hline
 Bus & $K_{en}$ [GWs] & $P_{out}$  & $K_{en}$ [GWs] & $P_{out}$ \\ 
 \hline
 1 & 34 & 9000 & 11.25 & 3000 \\ 
 \hline
 2 & 22.5 & 6000 & 22.5 & 6000 \\
 \hline
 3 & 7.5 & 2000 & 7.5 & 2000 \\
 \hline
 4 & 33 & 5000 & 20 & 3000 \\
 \hline
 5 & 13 & 2000 & 13 & 2000 \\
 \hline
\label{T:1}
\end{tabular}
\vspace{0.4cm}
\caption{Length of transmission lines.}
\begin{tabular}{c|c c c c c}
 \hline
 Lines & 1-2 & 1-4  & 2-3  & 2-4 & 3-5\\ 
 \hline
 Length [km] & 300 & 100  & 350 & 220 & 50\\ 
 \hline
 \label{T:2}
\end{tabular}
\end{table}

\subsection{Locational impact of frequency response from inverter-based resources}

In this part, the locational impact of frequency response from inverter-based resources is investigated. To evaluate the locational impact of frequency response, the disturbance response ratio $|R_{zd}(s)|$ is used as a metric. When $|R_{zd}(s)|<1$, then the control system is able to attenuate any external disturbance applied to the system. For this study, it is considered that the performance variable is the center of inertia frequency, the controlled inputs are the power outputs from the inverter-based resources at bus 1 and 3. Thus, two disturbance response ratios are defined namely $R_{zd}^1$ and $R_{zd}^3$. The inverter-based resources use local frequency measurements as feedback signal. The fast frequency controller considered for this analysis is depicted in Fig.~\ref{fig:Control_diagram}.

\subsubsection{High inertia case $K_{en}$=110 GWs}

\begin{figure}[t]
\centering
\includegraphics[width=1\linewidth, trim=3.8cm 8.1cm 4cm 8cm,clip]{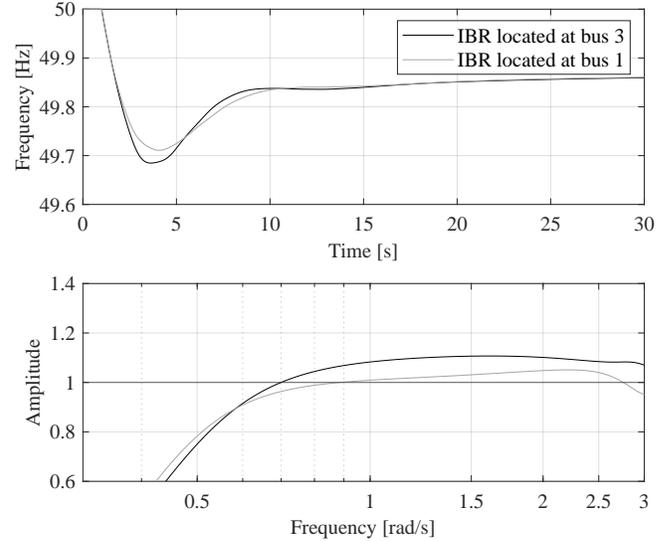}
\caption{Comparison of frequency response trajectory for different locations of the inverter-based resources. Disturbance response ratio to evaluate the locational impact of frequency response.}
\label{fig:DisturbanceRatioComparisonLocationIBR_bod}
\end{figure}
This study explores the behavior of the system for different locations of the inverter-based resources. A disturbance of 980 MW is applied at bus 2 in the system, which is the external disturbance to our system, $d_2=980 MW$ The total system inertia is equal to 110 GWs. Further, it is considered that 40\% of the total frequency response is provided by inverter-based resources, which can be located either at bus 1 or bus 3 and 60\% of the total droop resposne by hydro units.

Figure.~\ref{fig:DisturbanceRatioComparisonLocationIBR_bod} shows the comparison of frequency response trajectories and disturbance response ratios considering different locations of the inverter-based resources. It can be seen that when fast frequency reserves are located at bus 1 the maximum frequency deviation is lower compared to the case where primary frequency control is provided by the IBR located at bus 3 (49.7125 Hz compared to 49.686 Hz). Looking at the disturbance response ratios curves it can be seen that when fast frequency response is provided by bus 3 (black curve), the amplitude of the transfer function $|R_{zd}(s)|$ is higher than 1 within a larger frequency range (for $\omega>0.7$ rad/s) compared to the case where fast frequency response is provided by the IBR located at bus 1. This means that any disturbances within this range will be amplified and not attenuated. 

Moreover, when the fast frequency reserves are located at bus 1, the value of the crossover frequency of $R_{zd}$ (frequency at which $|R_{zd}|=1$) is equal to 0.87 rad/s and it is larger compared to the case where fast frequency response comes from the IBRs at bus 3. Thus, primary frequency control on bus 1 can better attenuate active power disturbances. This renders IBRs located at bus 1 more suitable for limiting the maximum frequency deviation in the system. Finally, since the frequency controller is droop-based, there is no contribution to the inertial frequency response, thus the rate of change of frequency is the same in both cases. Thus, it can be concluded when the inverter-based resources that provide fast frequency response are located close to the disturbance location, the frequency nadir of the center of inertia frequency has lower value.

\subsubsection{Low inertia case $K_{en}$=74.25 GWs}
The same comparison is conducted considering lower inertia level in the system. The total system inertia is equal to 74.25 GWs. Further, it is considered that 40\% of the total frequency response is provided by inverter-based resources, which can be located either at bus 1 or bus 3, and 60\% of the total droop response by hydro units.
\begin{figure}[t]
\centering
\includegraphics[width=1\linewidth, trim=3.8cm 8cm 4cm 8cm,clip]{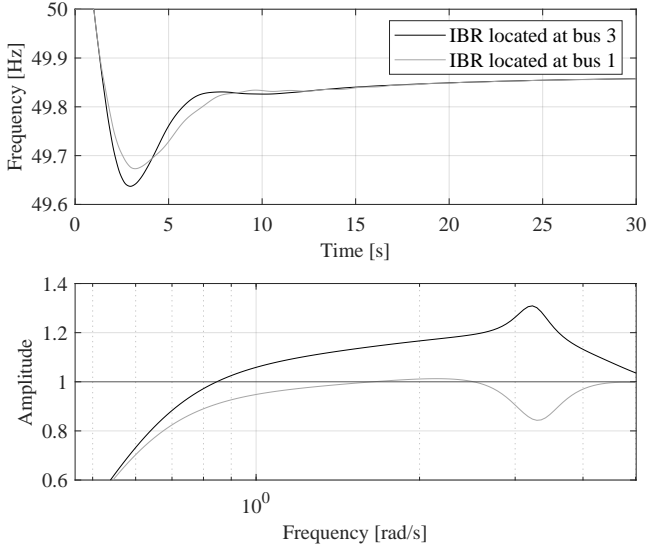}
\caption{Lower inertia scenario. Comparison of frequency response trajectory for different locations of the inverter-based resources. Disturbance response ratio to evaluate the locational impact of frequency response.}
\label{fig:DisturbanceRatioComparisonLocationIBR_bod_low_inertia}
\end{figure}

Figure.~\ref{fig:DisturbanceRatioComparisonLocationIBR_bod_low_inertia} shows the comparison of frequency response trajectories and disturbance response ratios considering different locations of the fast primary frequency control. It can be noticed that the reduction of system inertia influences the magnitude of $R_{zd}$ more, when primary frequency control is provided by active power regulating units located at bus 3. In particular, the maximum value of $R_{zd}$ is close to 1.3, which means that the frequency controller cannot attenuate well the active power disturbance occurring at bus 2. This is translated to higher maximum frequency deviation compared to the scenario with higher inertia level. Moreover, the region in which $|R_{zd}|$ is less than one, i.e. $|R_{zd}|<1$, is larger when fast frequency reserves are located at bus 1. This means that the maximum frequency deviation is smaller when fast frequency reserves are located at bus 1. This is confirmed from the time-domain analysis. when fast frequency reserves are located at bus 1 the maximum frequency deviation is lower compared to the case where frequency response is provided by the IBR located at bus 3 (49.6742 Hz compared to 49.6374 Hz). Notice that the difference between the frequency nadirs is higher for the lower-inertia scenario, since the difference in the amplitude of $R_{zd}$ increased considering the different fast frequency reserves' locations. This comes to show that the lower becomes the inertia level, the higher the locational impact becomes on the frequency response. Thus, to achieve the same frequency response and reduce the frequency nadir, system operators would have to procure more frequency reserves if the fast active power regulating units were located at bus 3.

\subsection{Impact of increased frequency droop response by inverter-based resources}

\begin{figure}[t]
\centering
\includegraphics[width=1\linewidth, trim=3.6cm 8cm 4cm 8cm,clip]{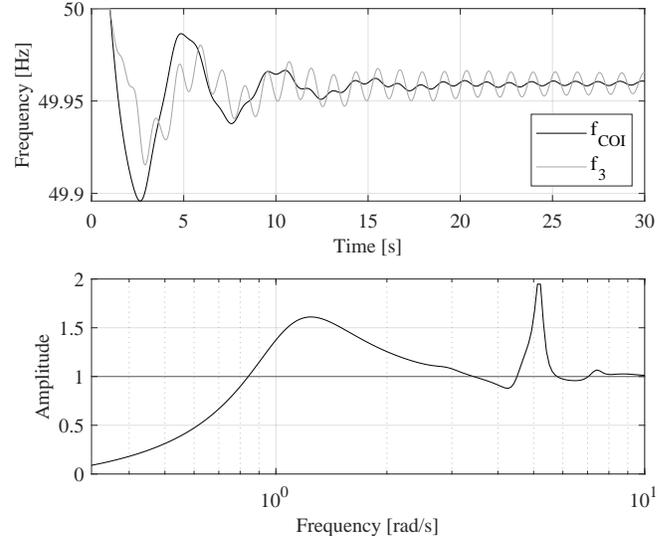}
\caption{System response to an increased total frequency droop dominated by inverter-based resources.}
\label{fig:IncreasedtotaldroopComparison}
\end{figure}
This study explores the behavior of the system for an increased droop response provided by inverter-based resources. A disturbance of 560 MW is applied at bus 2 in the system. The total system inertia is equal to 110 GWs. For this study the performance variable is the center of inertia frequency, i.e. $z=f_{\rm COI}$, and the measured output is the local frequency at bus 3, i.e. $y=f_{3}$. The external disturbance to the system $d$ is equal to the active power disturbance applied at bus 2. Moreover, 20\% of the primary frequency response is provided by the generator connected to bus 2 and the remaining 80\% by inverter-based resources connected to bus 3.

Figure~\ref{fig:IncreasedtotaldroopComparison} depicts the system response for increased total frequency droop and the response provided primarily by inverter-based resources and the amplitude of the disturbance response ratio. It can be seen that the peak of the sensitivity function is higher compared to all the previous scenarios, i.e. $|R_{zd}|_{\rm max}\approx2$. It can be inferred that for very large frequency droop response provided by active power regulating units at bus 3, that the frequency controller amplifies disturbances in regions where $|R_{zd}|>1$. This can cause excitation of local modes associated with electromechanical dynamics, since the peak of the sensitivity function is larger and consequently deteriorate the damping of the system. As it can be seen from the time-domain analysis, although the maximum frequency deviation is lower compared to the previous case, it can be noticed that frequency at bus 3 experiences oscillations of constant amplitude. Further increase of the frequency droop response by the inverter-based resources will result in small-signal instability of the system. It is interesting to notice that the frequency of oscillations is equal to 0.81 Hz, frequency at which $|R_{zd}|$ takes its maximum value.

Finally, this study highlights the fact that the stability margins might vary for different allocation of primary frequency control and the importance to evaluate the disturbance response ratio.

\subsection{Mitigation strategy}
Considering that the peak of $|R_{zd}|$ is considerably higher when fast frequency reserves are located at bus 3, one could limit its primary frequency contribution by setting a constraint on the maximum value of $|R_{zd}|$ and allocate primary frequency control at buses with smaller peak of $|R_{zd}|$. Consider the previous scenario, in which a disturbance of 560 MW applied at bus 2 in the system and the total system inertia is equal to 110 GWs. The objective is to allocate the 80\% of the total frequency droop control between buses 1 and 3 so that the maximum $|R_{zd}|$ reduces and the system becomes better damp. A maximum value of 1.35 is considered for the $R_{zd}$. The value 1.35 was empirically selected. Similarly to the previous cases, it is considered that the performance variable is the center of inertia frequency, the controlled inputs are the power outputs from the inverter-based resources at bus 1 and 3. Thus, two disturbance response ratios are defined namely $R_{zd}^1$ and $R_{zd}^3$. 

\begin{figure}[t]
\centering
\includegraphics[width=1\linewidth, trim=3.6cm 8cm 4cm 8cm,clip]{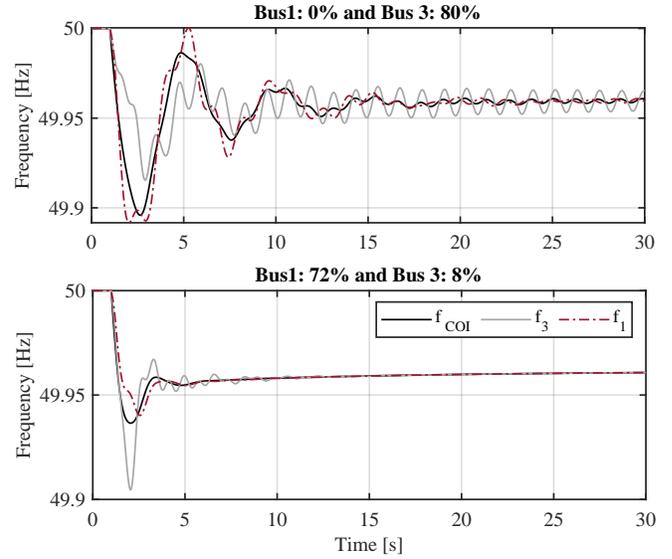}
\caption{Comparison of system response considering different allocation of fast primary frequency response between buses 1 and 3.}
\label{fig:IncreasedtotaldroopComparison_2}
\end{figure}

\begin{figure}[t]
\centering
\includegraphics[width=1\linewidth, trim=3.8cm 8cm 4cm 8cm,clip]{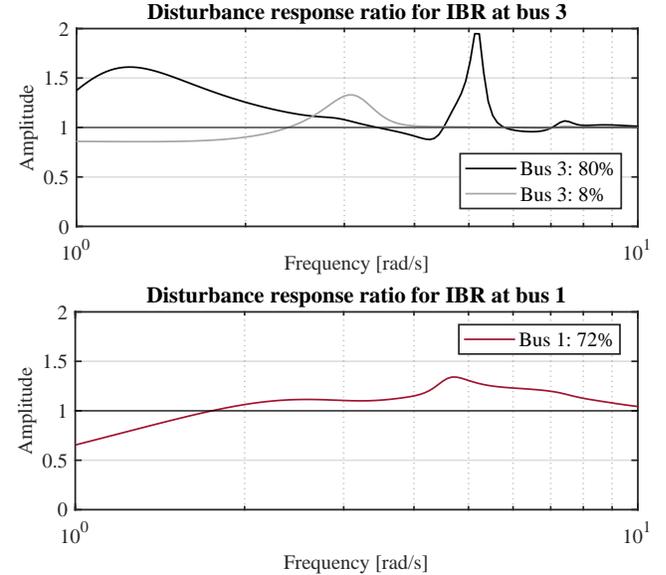}
\caption{Comparison of disturbance response ratios $R_{zd}^1$ and $R_{zd}^3$ considering different allocation of fast primary frequency response between buses 1 and 3.}
\label{fig:IncreasedtotaldroopComparison_bode}
\end{figure}

Figure~\ref{fig:IncreasedtotaldroopComparison_2} demonstrates the system response considering different allocation of fast primary frequency response between buses 1 and 3. We start from the case, in which 80\% of the total frequency droop control is provided by IBRs located at bus 3, which is similar to the previous scenario (80\% of total droop response is provided by IBRs and 20\% from generators). As shown in the figure, the system is marginally stable. Then, we gradually reduce the participation of the active power regulating units at bus 3 in primary frequency control until $R_{zd}^3$ becomes smaller than 1.35 (see Fig.~\ref{fig:IncreasedtotaldroopComparison_bode}). This results in 8\% of the total droop response being provided by IBRs at bus 3 and 72\% by IBRs at bus 1. It can be seen in Fig~\ref{fig:IncreasedtotaldroopComparison_2} that when 72\% of the total frequency droop response is provided by IBRs located at bus 1, the local frequencies $f_1$ and $f_3$, as well as the COI frequency, are better damped. This is expected since the maximum value for both disturbance response ratios $R_{zd}^1$ and $R_{zd}^3$ are lower than 1.35 (see Fig.~\ref{fig:IncreasedtotaldroopComparison_bode}). Thus, we realize that by allocating fast frequency response we can limit the maximum disturbance response ratios and avoid disturbance amplification and excitation of poorly damped modes in the system.

It should be mentioned that a more sophisticated approach, such as $H_{\infty}$/$H_2$ optimization or control design of a typical feedback damping controller, could be considered in order to achieve a more robust solution be achieved. However, this is not in the scope of this paper, therefore it is intended as future work.

\section{Conclusion}
As inertia level decreases and active power regulating units are located in remote areas, the nodal frequencies differ from the global one. This introduces control limitations for improving the frequency performance of the system, since the performance variables differ from the measured outputs. This paper assesses the impact of delivering fast primary frequency response. Using the disturbance response ratio of the system, the locational aspect of fast frequency reserves can be evaluated by means of screening the frequencies in which fast primary frequency control amplifies or attenuates disturbances in the system. The disturbance response ratio allows to evaluate how well the center of inertia frequency can be controlled by remotely placed active power regulating units. The results show that fast primary frequency reserves located at areas far from the disturbance location in combination with low-inertia scenarios can result in an unstable response due to excitation of poorly damped modes. In this case, careful consideration should be given on the selection of the total droop response provided by these reserves, as their increased frequency droop response can significantly deteriorate the small-signal stability margins of the system. A possible mitigation strategy for system operators would be to procure fast frequency reserves located in the electrical vicinity of disturbance location. Finally, our results show that fast primary frequency response is most effective in terms of reducing the frequency nadir and preserving the small-signal stability of the system, when it is provided from areas that are close to the location of disturbance. 

Future work will focus on utilizing the disturbance response ratio for allocating primary frequency response in large power systems and robustly designing primary frequency controllers in future power systems. 
\bibliographystyle{IEEEtran}
%



\bibliography{References}

\end{document}